\documentclass[journal,a4paper,]{IEEEtran}
\IEEEoverridecommandlockouts
\IEEEsettextheight{1in}{1.25in}
\IEEEsettopmargin{a}{0.25in}
\IEEEsettextwidth{19mm}{19mm}
\IEEEsetsidemargin{i}{19mm}

\usepackage{cite}
\usepackage{amsmath,amssymb,amsfonts}
\usepackage{algorithmic}
\usepackage{graphicx, subfig}
\usepackage{textcomp}
\usepackage{xcolor}
\usepackage{multirow}
\usepackage{tfrupee} 
\def\BibTeX{{\rm B\kern-.05em{\sc i\kern-.025em b}\kern-.08em
    T\kern-.1667em\lower.7ex\hbox{E}\kern-.125emX}}

\begin{document}

\title{Simplified Algorithm for Dynamic Demand Response in Smart Homes Under \\ Smart Grid Environment
\thanks{This publication is an outcome of the R\&D work undertaken in the project under the Visvesvaraya PhD Scheme of Ministry of Electronics and Information Technology, Government of India, being implemented by Digital India Corporation (formerly Media Lab Asia).}
}
\author{Shashank~Singh,
		Aryesh Namboodiri,
        and~Selvan~M.P.\\
        \textit{Hybrid Electrical Systems Laboratory, Department of Electrical and Electronics Engineering}
        \\\textit{National Institute of Technology Tiruchirappalli, Tamil Nadu 620015, India}
        \\shashanksingh0110@gmail.com, aryeshvn@gmail.com, selvanmp@nitt.edu}

\markboth{Presented in 2019 IEEE PES GTD Grand International Conference and Exposition Asia (GTD Asia), Bangkok, Thailand}{Shell \MakeLowercase{\text
it{et al.}}: A Novel Tin Can Link}
\maketitle

\begin{abstract}

Under Smart Grid environment, the consumers may respond to incentive--based smart energy tariffs for a particular consumption pattern. Demand Response (DR) is a portfolio of signaling schemes from the utility to the consumers for load shifting/shedding with a given deadline. The signaling schemes include Time--of--Use (ToU) pricing, Maximum Demand Limit (MDL) signals etc. This paper proposes a DR algorithm which schedules the operation of home appliances/loads through a minimization problem. The category of loads and their operational timings in a day have been considered as the operational parameters of the system. These operational parameters determine the dynamic priority of a load, which is an intermediate step of this algorithm. The ToU pricing, MDL signals, and the dynamic priority of loads are the constraints in this formulated minimization problem, which yields an optimal schedule of operation for each participating load within the consumer provided duration. The objective is to flatten the daily load curve of a smart home by distributing the operation of its appliances in possible low--price intervals without violating the MDL constraint. This proposed algorithm is simulated in MATLAB environment against various test cases. The obtained results are plotted to depict significant monetary savings and flattened load curves.  
\end{abstract}
\vspace*{.5em}
\renewcommand\IEEEkeywordsname{Keywords}
\begin{IEEEkeywords}
Demand response, dynamic priority, load scheduling, smart grids, time of use (ToU) pricing
\end{IEEEkeywords}

\section{Introduction}
\IEEEPARstart{T}{he} integration of renewable energy sources, information and communication technology (ICT), compliance with Industry 4.0 \cite{R1} are transforming legacy grid into smart grid (SG) \cite{R2}. SG facilitates energy management, monitoring and sophisticated control schemes through the integration of advanced metering infrastructures, ICT, smart energy meters, smart appliances/loads, micro \& distributed generations, and electricity storage techniques. Further, SG also opens the access to electricity markets through increased transmission paths, aggregated supply and demand response initiatives, and ancillary service provisions  \cite{R3, singh2018}.

Demand Response (DR) is one of the distinguished features of SG. DR programs are the attempts which encourage consumers to alter their consumption pattern according to the incentive based signaling schemes rather than adjusting generation levels. This helps in maintaining a healthy balance between the electrical power demand and supply through the techniques like load shifting and load clipping during peak load or peak pricing hours \cite{R7}. 

The electric utility company (the utility) may implement DR by introducing seasonal energy tariffs, concessional rates for the energy consumption in off-peak hours \cite{R8}. Conventionally, the utilities have adopted the flat rate of electricity tariff structure, wherein the consumers are charged at fixed price of electricity per unit throughout the day. Another type of pricing scheme is dynamic pricing which involves time varying or time--of--use (ToU) price of electricity over a constant interval throughout a day, which makes this scheme suitable for DR. Reference \cite{R9} presents the importance of the dynamic pricing over the flat rate structure. Further, it has emphasized the need for re-framing the regulations and policies of the current electricity market to supply electricity to the consumers at its true cost. Reference \cite{R10} has suggested another signaling scheme, known as power import limit (PIL), to implement DR. The PIL refers to an upper limit which defines a maximum demand limit (MDL); i.e. the maximum amount of electric power which can be drawn from the grid during an interval without additional penalty.

The term smart home uses automated processes to control the building's operations including heating, ventilation, air conditioning, lighting, security, and other systems. It uses sensors, actuators, and microchips, in order to collect data and manage it according to business functions and services. This infrastructure optimizes the net energy consumption, efficiently utilizes space to have minimal impact on environment. The ICT enables smart homes to communicate with both its internal devices, appliances (which it can control), and also with its surroundings. Furthermore, a smart home can adapt to grid's conditions and communicate with neighbour smart homes, thereby creating an active or a virtual power plant \cite{R12, R13, R14, marzband2018a}. 

In this paper, a minimization problem has been formulated to showcase the DR in a smart home. There are several algorithms available in the literature which deal with similar optimization problems \cite{R15, marzband2018b}. A heuristic--based evolutionary algorithm is proposed in \cite{R16} that easily adapts heuristics in the load scheduling problem for achieving substantial saving on energy costs, and thus reducing the peak energy demand for residential, commercial and industrial consumers. Ali \textit{et al.} \cite{R17} have formulated a load scheduling algorithm, which employs the two--component based ToU tariff structure to obtain substantial reduction in consumer's electricity bill without forcing utility to add more generation to total installed capacity. Furthermore, heuristic based algorithms for energy management systems are introduced. Their course of action includes reduction of load during peak demand \cite{R18}, establishing trade--off among the electricity bill and the waiting time for the operation of every household appliance \cite{R19}, learning automata based scheduling \cite{R20}, reward based DR \cite{R8}, and real--time pricing based DR for electric vehicles \cite{R21}. Researches have also implemented DR among multiple smart homes, wherein an energy trading and information exchange framework among the multiple smart homes with distributed generation facility has been developed in \cite{marzband2018d, marzband2018e, marzband2018c}. 

The aforementioned techniques/frameworks have not dealt with the dynamic priority, user input, and utility signals for effective load scheduling. Further, neither the operational dynamics of the nonschedulable loads nor the influence of higher operational priority load on next priority load have been considered. Another impediment is the estimation of the impact of penalty incurred if the noninterruptible loads are in ON condition and suddenly user turns ON the other noninterruptible load \cite{R10}, thus violating PIL constraint. This kind of real--time situations have been bottleneck for aforementioned algorithms. Considering the limitations of previous works, a new residential load--scheduling algorithm has been developed and tested in this paper. The developed algorithm is based on the dynamic priority of the loads, and bounded by ToU tariff and PIL constraints. The major contributions of this paper are as follows:
\begin{itemize}
\item Concept of dynamic priority based DR is applied where the algorithm considers user defined operational load parameters to compute the dynamic priority of a load for its scheduling. 
\item The DR algorithm is bounded by two constraints; i.e. PIL and ToU pricing, and aimed to yield substantial savings in the energy bills for an individual consumer. 
\item The dynamics of thermostatically controlled appliances (TCA) (e.g. air--conditioners, water heaters) are also considered.
\item A provision to extend the temperature limits of TCAs is introduced, which is referred here as tolerance. The tolerance enables effective scheduling of other appliances during the turn OFF timings of the TCAs. 
\end{itemize}
The aforementioned claims are simulated against various test cases considering variety of residential loads in MATLAB. The results are plotted as well as tabulated to showcase the efficacy of the algorithm. The structure of this paper is as follows: Section--II deals with problem formulation, wherein load classification, mathematical formulation of objective function, its input and constraints are considered. The solution methodology and its intermediate steps are described in section--III. Section--IV contains the simulated results, comparison and benefits of the presented algorithm. The paper is concluded in section--V.

\section{Problem Formulation}
In this paper, the residential load scheduling problem is dealt with a function which computes the dynamic priority of loads in real--time and schedules their operation under precursory constraints. The precursory operational constraints and parameters involved in this problem must be listed, framed and analyzed properly to arrive at an optimal solution. 

The primary operational constraint is the category of domestic appliances. Based on the scope of applications, the domestic appliances are classified into four categories, namely, noninterruptible--nonschedulable loads (NINSLs), interruptible--nonschedulable loads (INSLs), noninterruptible--schedulable loads (NISLs), and interruptible--schedulable loads (ISLs) as described in \cite{R10}. The NINSLs and INSLs can be grouped into nonschedulable loads (NSLs), whereas the NISLs and ISLs can be grouped into schedulable loads (SLs).

\subsection{Scheduling Interval}
Primarily, the algorithm considers scheduling of the SLs without influencing the operation of NINSLs. It monitors the aggregate power consumption of participating loads at real--time to schedule the SLs. The span of one day (24 hours) is divided into several scheduling intervals to incorporate the dynamic changes occurring from the utility side or at the consumer side. The duration of scheduling interval is assumed to be the minimum run--time of any appliance. Further, the price of electricity and PIL are assumed to be constant during a scheduling interval. For simplicity, an interval having duration of $5$ minutes is considered as the duration of scheduling interval in this paper, which accounts for $288$ such intervals in a day ranging from $1^{st}$ interval ($00:00:00-–00:04:59$) to $288^{th}$ interval ($23:55:00–-23:55:59$).
\subsection{Input Specification}
There must be sufficient input parameters, e.g. the electricity tariff scheme, the particulars about the SLs available at the consumer premises for the formulation of a DR algorithm. Since the operational timings of NINSLs entirely depend upon the desire and comfort of the consumer, the algorithm does not incorporate their operational timings. But the power consumption by NINSLs are considered in this algorithm. For INSLs, the set point temperature and desired tolerance values must be provided prior to its operation. The required inputs are compiled as follows:
\subsubsection{Electricity Tariff}
In this paper, two types of tariff structures have been considered, i.e. Flat rate tariff and ToU tariff. In Flat rate tariff, the price of electricity--per--unit remains constant irrespective of time of day, season, etc. In ToU tariff, the utility communicates the time--varying electricity price per unit to consumer well in advance, mostly one day before. The utility pricing--interval is considered to be of $60$ minutes. 
\subsubsection{Power Import Limit (PIL)}
Similar to ToU tariff, the utility notifies an upper limit on power import to the consumers well in advance. Here, two cases of PIL are considered namely fixed and dynamic. Fixed PIL does not vary during the day, while the dynamic PIL is expected to vary at an interval of $60$ minutes. The considered penalty is twice the normal price of electricity per--excess--unit for that particular interval.
\subsubsection{Consumer Load Type and Ratings}
The consumers may define which of their loads are schedulable and nonschedulable. The algorithm requires rated power specifications of the all participating loads (including NINSLs and INSLs).
\subsubsection{Specifications}
For any SLs, the consumers must provide the three timing parameters such as start time interval ($s$), stop time interval ($f$, interval by which the load must complete its operation), and run time interval ($r$, duration of operation of load in minutes). The INSLs include heating loads and cooling loads. The consumer must specify the type of INSL, its set point temperature, and tolerance.

\subsection{Objective Function}
Let's say $n$ be the total number of SLs connected at user premises, then the residential load scheduling algorithm can be expressed as a minimization problem using \eqref{eq1}. 
\begin{equation}
F\left(j,k\right)=\sum_{j=1}^{n}\sum_{k=1}^{288}\left(C^k \times P_j \times u_j^k\right)
\label{eq1}
\end{equation}
where, $C^k$ denotes the price of energy per $kWh$ during $k^{th}$ scheduling interval. $P_j$ denotes the power rating of $j^{th}$ load. The term $u_j^k$ denotes the operating status of $j^{th}$ load during $k^{th}$ interval. $u_j^k \leftarrow (1/0)$ represents the load's binary operating status, i.e. \textbf{On/Off}. $F\left(j,k\right)$ represents the total electricity bill of a consumer. The operational constraints to minimize \eqref{eq1} are as follows: 

\textit{(a)} The total running intervals of any load $j$ must be equal to its specified $r_j$ as shown in \eqref{eq2}. \textit{(b)} The operating status of $j^{th}$ load can be represented using \eqref{eq3}. \textit{(c)} At an interval $k$, the net power consumption must be within ${PIL}^k$ otherwise penalty of twice the regular price of electricity is charged to the consumer for excess consumption \eqref{eq4}. \textit{(d)} For each INSL (say, air--conditioner (AC)), the algorithm should enable INSL to maintain the operating room temperature $(T_j^k)$ based on set point temperature $(T_j^{set})$ and tolerance $(\Delta T_j)$ at interval $k$. The operational constraints of cooling type INSL is expressed using \eqref{eq5}. 
\begin{equation}
\sum_{k=1}^{288}{u_j^k} \times 5 = r_j
\label{eq2}
\end{equation}
\begin{equation}
u_j^k=0;~\textrm{for}~k\ >f_i~\textrm{and}~k<s_j
\label{eq3}
\end{equation}
\begin{equation}
\sum_{j=1}^{n}{P_j \times u_j^k\le{PIL}^k}
\label{eq4}
\end{equation}
\begin{equation}
u_{j}^{k} = \left\lbrace
\begin{matrix}
   0 & T_{room}^{k-1}<T_{j}^{set}  \\
   1 & T_{room}^{k-1}>T_{j}^{set}+\Delta {{T}_{j}}  \\
   {u}^{k-1} & T_{j}^{set} \le T_{room}^{k}\le T_{j}^{set}+\Delta {{T}_{j}}  \\
\end{matrix}
\right.
\label{eq5}
\end{equation}

\section{Solution Methodology}
First, the consumer enters the operational specifications of $j^{th}$ SL at least one interval prior to its $s_j$. In the beginning of each $k^{th}$ interval, the algorithm records the power consumption of the NINSLs and INSLs before computing the $r_j$ of each $j^{th}$ SL to compute the available ${PIL}^k$. Further, the power consumption of NSLs, if already turned ON, is assumed to be constant throughout that particular interval. For any interval $k$ having $m$ number of NSLs, the ${PIL}^k$ can be updated using \eqref{eq6}.
\begin{equation}
{\rm PIL}_{new}^k={\rm PIL}_{old}^k-\sum_{j=1}^{m}P_{{\rm NSL}_j}^k
\label{eq6}
\end{equation}
where, ${\rm PIL}_{new}^k$ denotes the updated $PIL$ for an interval $k$. ${\rm PIL}_{old}^k$ denotes the utility specified $PIL$ for $k^{th}$ interval. ${\rm P}_{{NSL}_j}^k$ denotes the power consumption of $j^{th}$ NSL during $k^{th}$ interval. Thereafter, the dynamic priorities ($DP$) of SLs are computed using \eqref{eq7} for their scheduling.
\begin{equation}
DP =\frac{Current~Interval}{Stop~Time~Interval-Run~Time~Interval+1}
\label{eq7}
\end{equation}

Here, $DP$ ranges from $0 \rightarrow 1$. If for a particular load, $DP \gets 1$ \textbf{or} $r_j == f_j - k$, then the particular $j^{th}$ load must turn ON and will remain ON until it completes its scheduled run time irrespective of any constraint. The loads having $DP < 1$, are considered for scheduling based on the highest $DP$ and available $PIL$ in $k^{th}$ interval.

The minimum cost of operation of any load during an interval $k$ can only be achieved if the price of electricity in that interval is low and $PIL$ is high. This leads to introduction of a ratio, referred here as Cost--to--Power ($CPR$) ratio. The $CPR$ for $j^{th}$ load in $k^{th}$ interval is expressed using \eqref{eq8}.
\begin{equation}
{CPR}_j^k=\frac{P_j \times C^k \times r_j}{60 \times {PIL}^k}
\label{eq8}
\end{equation}

\begin{figure}[htbp]
\centering
\subfloat[Flow chart of load scheduling algorithm]{\includegraphics[width=\columnwidth]{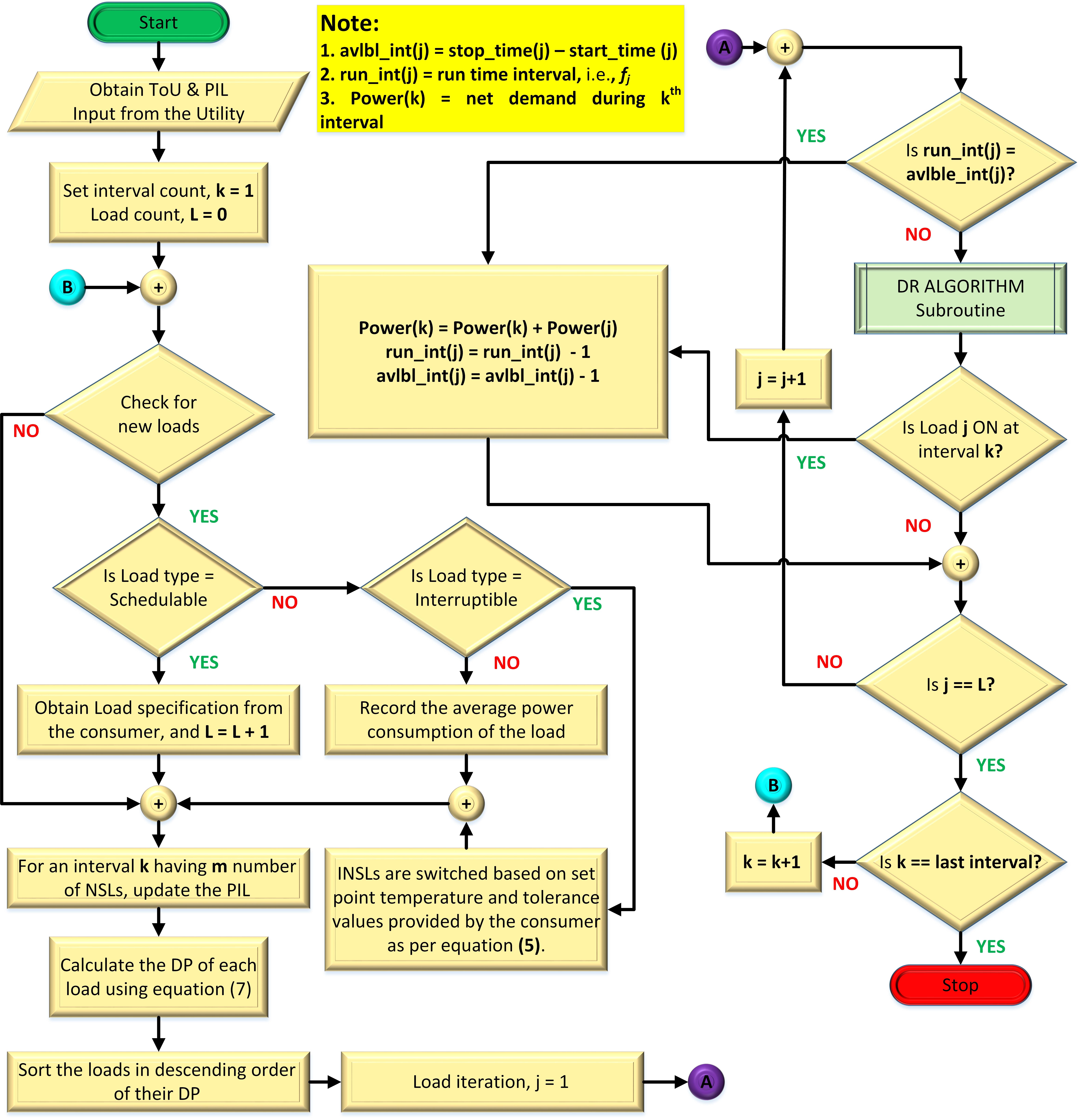}\label{Fullalgo}}
\hfill
\centering
\subfloat[DR algorithm subroutine]{\includegraphics[width=\columnwidth]{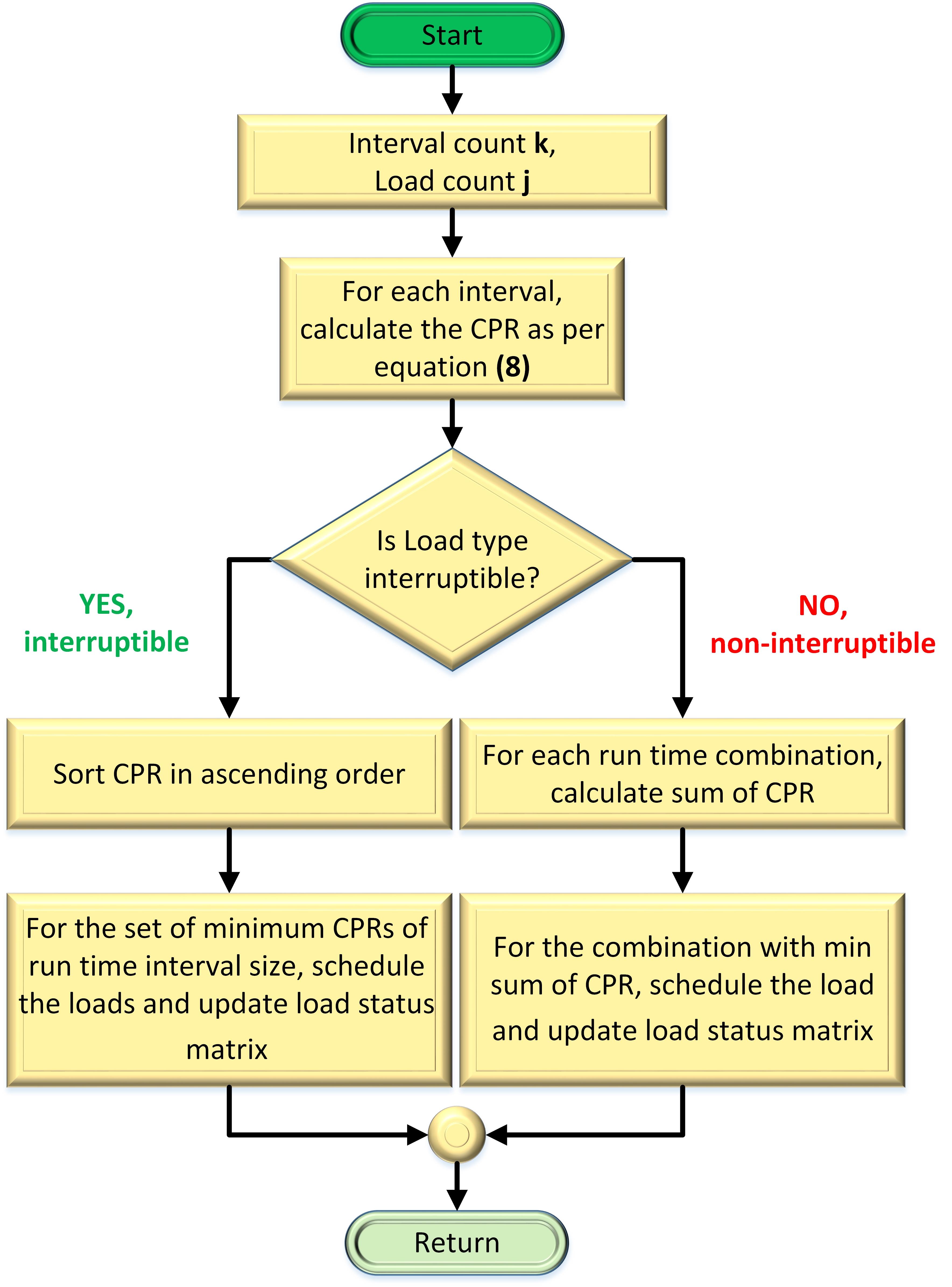}\label{DRalgo}}
\caption{Solution of formulated minimization problem}
\label{fig1}
\end{figure}

${CPR}_j^k$ of the $j^{th}$ load is dynamically computed for each $k^{th}$ interval between load's $s_j$ and $f_j$. The set of intervals having minimum $CPR$ is the optimal solution, thereby a particular SL is scheduled in those intervals. The intermediate steps of this algorithm is illustrated through flowchart in Fig.~\ref{fig1}, wherein, the subroutine from Fig.~\ref{Fullalgo} is depicted in Fig~\ref{DRalgo}. 

\section{Case Study and Results}
\begin{figure}
\centering
\includegraphics[scale=0.8]{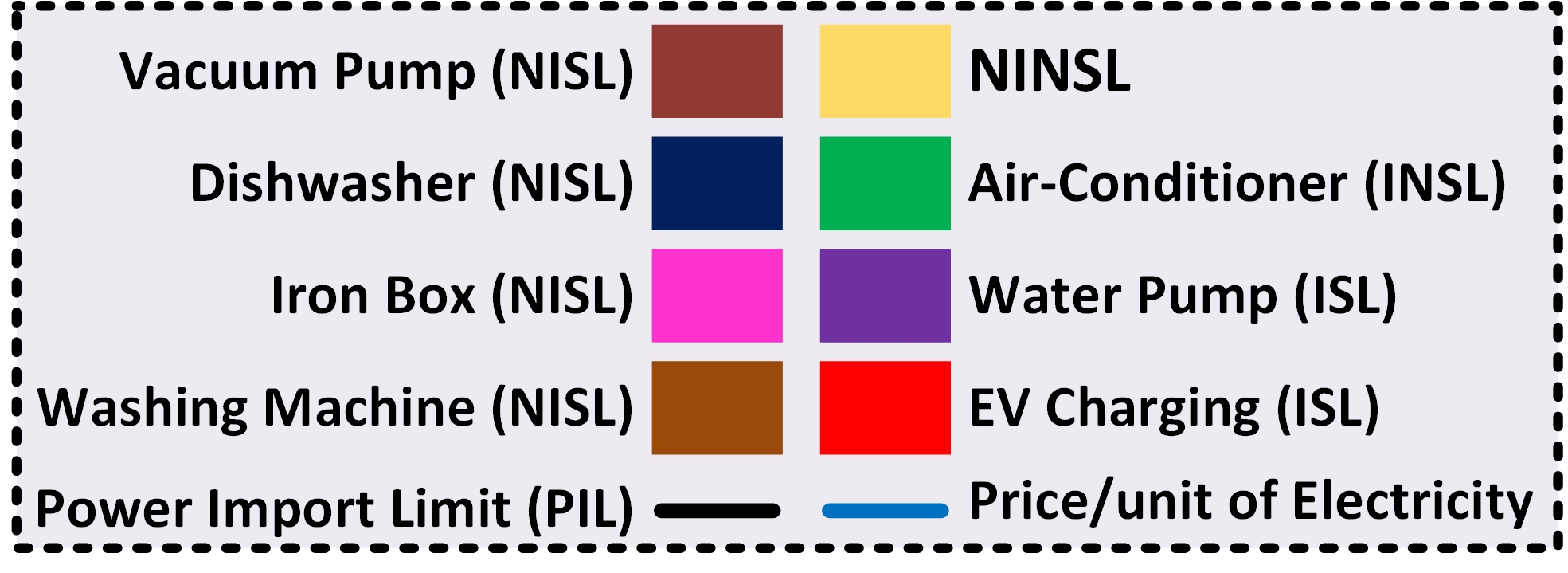}\
\caption{Colors used to illustrate the time of operation loads}
\label{colours}
\end{figure}

\begin{figure}[htbp]
\centering
\includegraphics[scale=0.5]{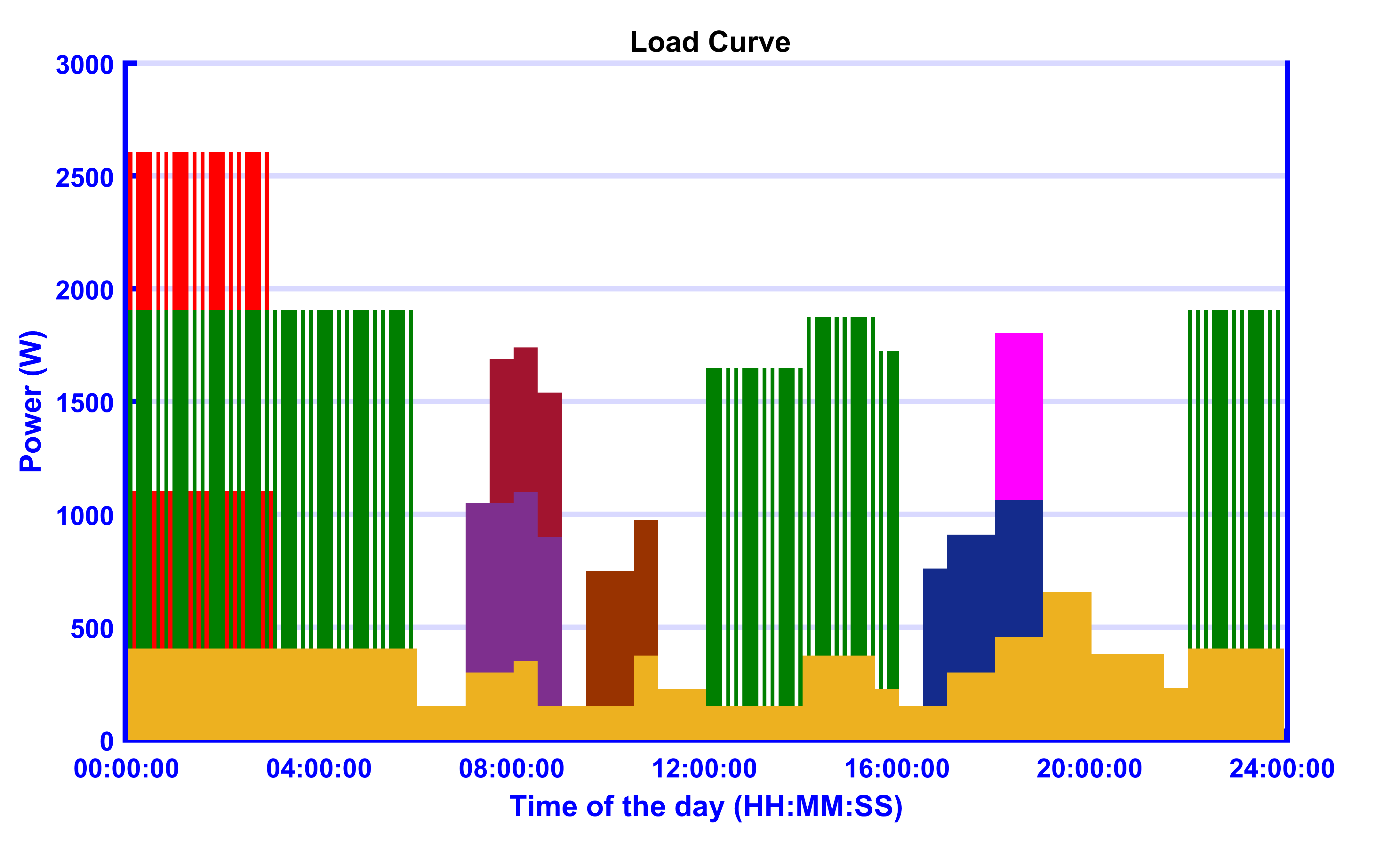}
\caption{Illustration of manual operation of loads}
\label{NormalOperation}
\end{figure}

Considering the utility provided signaling schemes, three case studies have been formulated for the same consumer behavior. The efficacy of the algorithm is presented through the simulation studies in MATLAB environment. To justify the comparison and benefits, the ratings of various participating NINSLs, INSLs and SLs, operating time specifications, ToU pricing, and dynamic PIL must be same. Hence, the rating of various SLs, NINSLs, and INSLs are described in Table~\ref{LoadsDescription}. Their operating time specifications are listed in Table~\ref{InputParameters}. Further, the color codes representing the duration of operation of a particular home appliance are shown in Fig~\ref{colours}. Fig.~\ref{NormalOperation} depicts the switching of participating loads as per the consumer's choice (Table~\ref{InputParameters}). In this case no scheduling algorithm is applied. The obtained behavior is compared with the proposed scheduling algorithm through forthcoming three cases to analyze the monetary savings in energy consumption.
\begin{table*}[htbp]
\renewcommand{\arraystretch}{1.1}
\caption{Description of Various Participating Loads}
\centering
\begin{tabular}{c c c c | c c c | c c c}
\hline
\hline
  & \multicolumn{3}{c|}{SLs} & \multicolumn{3}{c|}{NINSLs} & \multicolumn{3}{c}{INSLs}\\
\hline
 \textbf{S. No.} & \textbf{Loads} & \textbf{Qty} & \textbf{Rating (W)} & \textbf{Loads} & \textbf{Qty} &\textbf{Rating (W)} & \textbf{Loads} & \textbf{Qty} &\textbf{Rating (W)} \\
\hline
1. &	Water Pump (ISL) &	1 &	750 & Ceiling Fan & 3 & 75 each & \multirow{6}{*}{Air-conditioner} & \multirow{6}{*}{1} & \multirow{6}{*}{1500} \\
2. &	Washing Machine (NISL) & 1 & 600 & Lighting Loads & 5 &	80 aggr. & & & \\
3. &	Vacuum Cleaner (NISL) & 1 & 640 & Refrigerator &	1 &	150 & & & \\
4. &	Dishwasher (NISL) &	1 & 610 & Bed Lamp &	2 &	10 each & & & \\
5. &	Iron Box (NISL)	& 1 & 740 & Computer &	1 &	200 & & & \\
6. &	EV Charging (ISL) & 1 & 700 & Television & 1 & 150 & & & \\
\hline
\label{LoadsDescription}
\end{tabular}
\end{table*}
\begin{table}[htbp]
\renewcommand{\arraystretch}{1}
\caption{Input Parameters}
\centering
\setlength\tabcolsep{4pt} 
\begin{tabular}{c c c c c c}
\hline
\hline
\textbf{S.No.} & \textbf{Load Name} & $s$ & $f$ & $r$ & $OT$\\
\hline
1. & Water Pump & 07:00 &	10:30 &	02:00 & 07:00 -- 09:00 \\
2. & Washing Machine & 09:30 &	12:00 &	01:30 & 09:30 -- 11:00 \\
3. & Vacuum Cleaner & 07:30 &	10:00 &	01:30 &	07:30 -- 09:00 \\
4. & Dishwasher & 16:30 &	20:00 &	02:30 &	16:30 -- 19:00 \\ 
5. & Iron Box & 18:00 &	21:00 &	01:00 &	18:00 -- 19:00 \\
6. & Electric Vehicle & 00:00 & 07:00 & 03:00 & 00:00 -- 03:00 \\
\multirow{3}{*}{7} & \multirow{3}{*}{Air-Conditioner} &  &  &  & 00:00 -- 06:00 \\
					&					&  &  &  & 12:00 -- 16:00\\
					&					&  &  &  & 22:00 -- 00:00\\
\hline
\multicolumn{6}{c}{$s$, $f$, $r$, and $OT$ (operating time) are represented in $HH:MM$}
\label{InputParameters}
\end{tabular}
\end{table}

\subsection{Case 1: With Dynamic PIL and Flat Rate Tariff}
The total electricity bill is analyzed under PIL, which varies hourly in a day. The price per unit of electricity is considered to be constant throughout the day. Here, the SLs are switched as per the schedule dictated by the proposed algorithm. The load curve obtained from this case is depicted in Fig.~\ref{case1b}. 

\subsection{Case 2: With Static PIL and ToU Tariff}
The total electricity bill is analyzed under static PIL, which remains constant throughout the day. The ToU tariff is considered, wherein the price per unit of electricity varies hourly in a day. Similar to case 1, the switching of loads according to proposed algorithm is shown in Fig.~\ref{case2b}. 

\subsection{Case 3: With Dynamic PIL and ToU Tariff}
\begin{figure}
\centering
\subfloat[Case 1]{\includegraphics[scale=0.35]{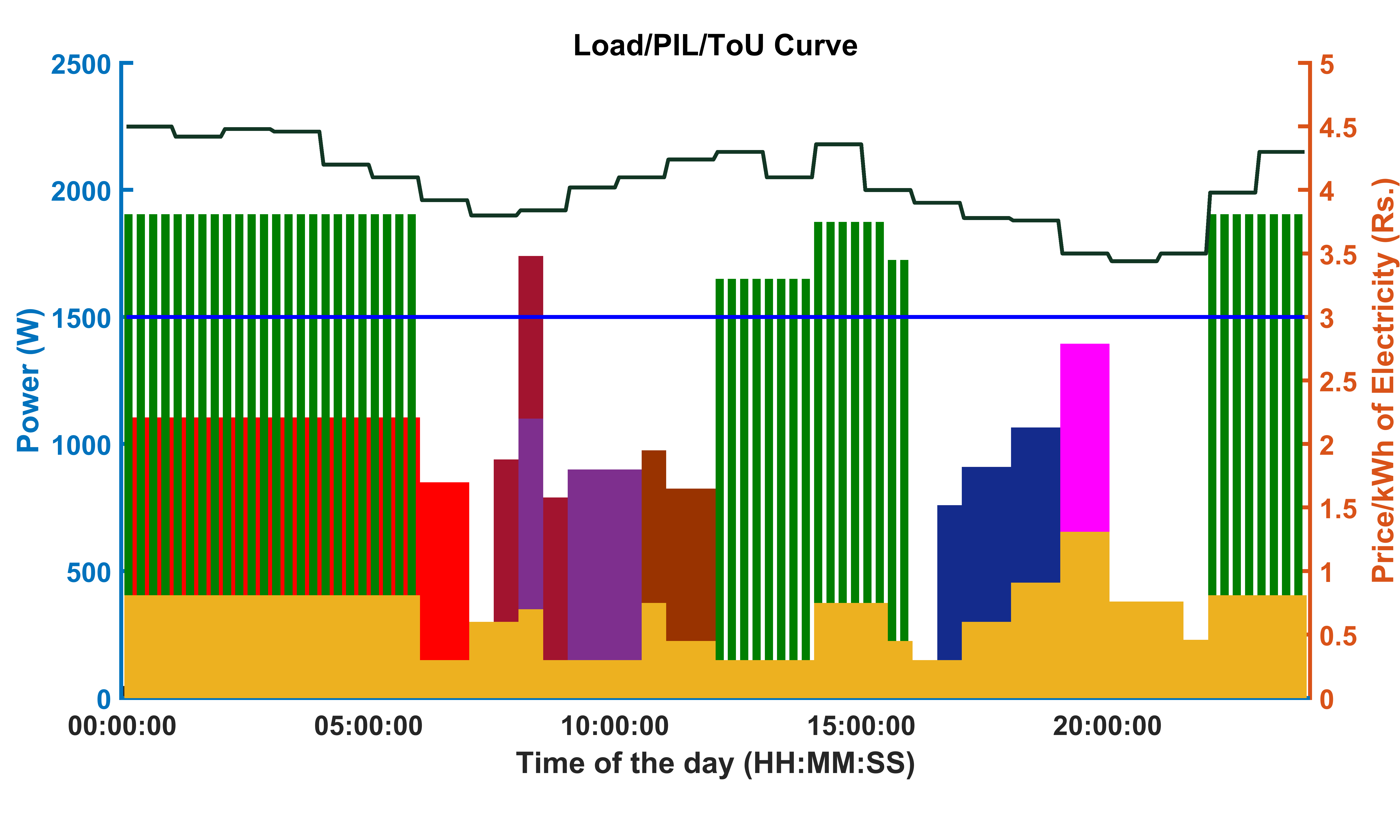}\label{case1b}}
\hfill
\subfloat[Case 2]{\includegraphics[scale=0.35]{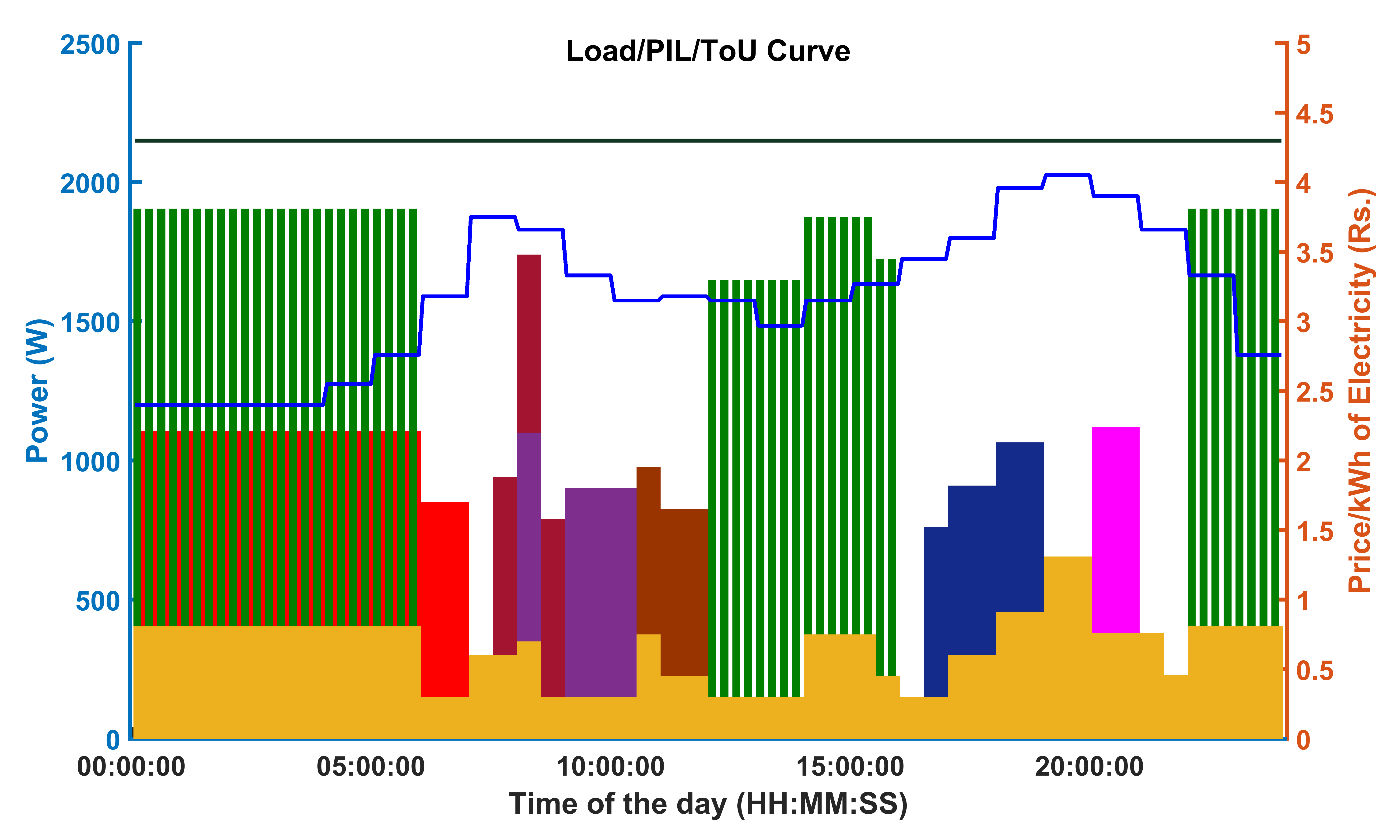}\label{case2b}}
\hfill
\subfloat[Case 3]{\includegraphics[scale=0.35]{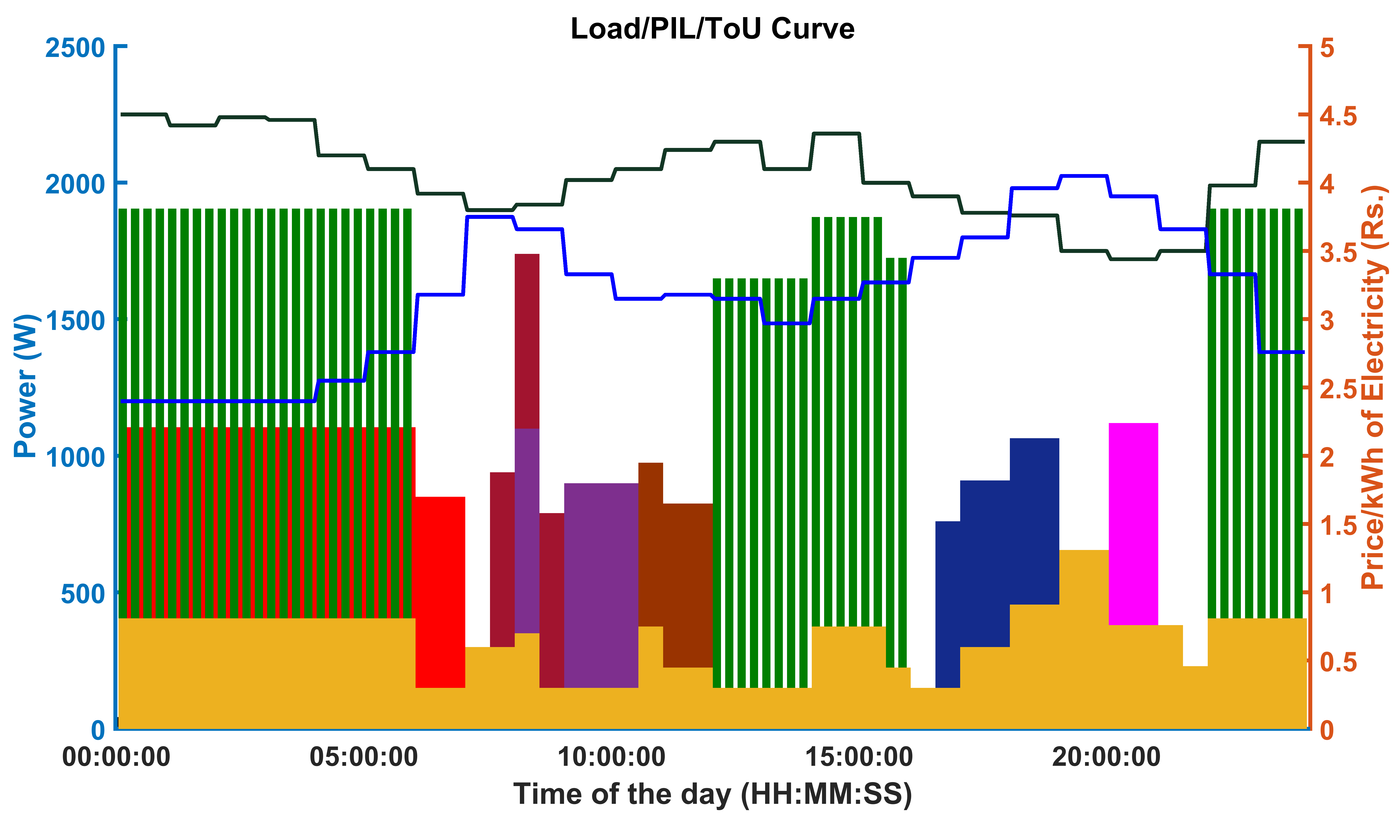}\label{case3b}}
\hfill
\caption{Scheduled operation by DR algorithm}
\label{case123}
\end{figure}

The total electricity bill is analyzed under dynamic PIL and ToU tariff. Similar to case 1 and 2, the switching of loads according to proposed algorithm is shown in Fig.~\ref{case3b}. 

Finally, for all three cases, the scheduled operating time of loads and cost of operation are tabulated in Table~\ref{AllCases}. It can be observed from the Fig.~\ref{NormalOperation} that the electric vehicle (EV) charging and AC have been operated together. Due to that, during ON period of AC, the net-power consumption has crossed the PIL and penalty charge is incurred. Whereas after the implementation of DR algorithm, the EV charging is interrupted, and turned ON only when AC is in OFF condition, which thereby eliminates the penalty cost. The Water Pump is scheduled with interrupted operation at multiple intervals to have optimal cost of operation. The Iron Box is also scheduled in low price intervals. Similar operating behavior of loads are reported in case 2 and 3. The savings on energy bills is about 3.97\% and a reduction in peak demand of about 30\% can be inferred from case 1. During case 2, the net savings is only due to elimination of penalty component which accounts for 5.24\% savings in the energy bill and 30\% reduction in peak load. In case 3, 4.9\% savings in energy bill and 30\% reduction in peak demand is observed. Furthermore, relatively flat load profiles have resulted due to proposed DR algorithm. The findings are reported in Table~\ref{AllCases} and illustrated in Fig.~\ref{case123}, which justifies the efficacy of the algorithm.

\begin{table*}[]
\caption{Comparison of operating time of participating loads under various cases}
\centering
\setlength\tabcolsep{4pt} 
\begin{tabular}{|c|c|c|c|c|c|c|c|c|c|c|}
\hline
\multirow{2}{*}{\textbf{Load Name}} & \multirow{2}{*}{\textbf{OT}}   & \multicolumn{3}{c|}{\textbf{Case-1}}                                & \multicolumn{3}{c|}{\textbf{Case-2}}                                & \multicolumn{3}{c|}{\textbf{Case-3}}                                \\ \cline{3-11} 
                                    &                                & \textbf{RC}           & \textbf{SRT}        & \textbf{ORC}           & \textbf{RC}           & \textbf{SRT}        & \textbf{ORC}           & \textbf{RC}           & \textbf{SRT}        & \textbf{ORC}           \\ \hline
\multirow{2}{*}{Water Pump}         & \multirow{2}{*}{07:00 -- 09:00} & \multirow{2}{*}{5.56} & 08:00 -- 08:30         & \multirow{2}{*}{5.05} & \multirow{2}{*}{4.50} & 08:00 -- 08:30         & \multirow{2}{*}{4.50} & \multirow{2}{*}{5.55} & 08:00 -- 08:30         & \multirow{2}{*}{5.05} \\ \cline{4-4} \cline{7-7} \cline{10-10}
                                    &                                &                       & 09:00 -- 10:30         &                       &                       & 09:00 -- 10:30         &                       &                       & 09:00 -- 10:30         &                       \\ \hline
Washing Machine                     & 09:30 -- 11:00                  & 2.89                  & 10:30 -- 12:00         & 2.85                  & 2.70                  & 10:30 -- 12:00         & 2.70                  & 2.89                  & 10:30 -- 12:00         & 2.85                  \\ \hline
Vacuum Cleaner                      & 07:30 -- 09:00                  & 3.54                  & 07:30 -- 09:00         & 3.54                  & 2.88                  & 07:30 -- 09:00         & 2.88                  & 3.54                  & 07:30 -- 09:00         & 3.54                  \\ \hline
Dishwasher                          & 16:30 -- 19:00                  & 5.66                  & 16:30 -- 19:00         & 5.66                  & 4.58                  & 16:30 -- 19:00         & 4.58                  & 5.66                  & 16:30 -- 19:00         & 5.66                  \\ \hline
Iron Box                            & 18:00 -- 19:00                  & 2.93                  & 20:00 -- 21:00         & 2.88                  & 2.22                  & 19:00 -- 20:00         & 2.22                  & 2.93                  & 19:00 -- 20:00         & 2.88                  \\ \hline
EV Charging                         & 00:00 -- 03:00                  & 8.62                  & M.I.* & 5.70                  & 10.78                 & M.I.* & 6.30                  & 9.41                  & M.I.* & 5.70                  \\ \hline
NINSLs                         & --                 & 23.39                  & -- & 23.39                  & 24.79                 & -- & 24.79                  & 24.79                  & -- & 24.79                  \\ \hline
INSL (AC)                         & --                  & 36.00                  & -- & 36.00                  & 33.57                 & -- & 33.54                  & 33.57                  & -- & 33.54                  \\ \hline
\multicolumn{2}{|c|}{\textbf{Total Cost (\rupee~)}}                                     & \textbf{88.59}                  &                     & \textbf{85.07}                 & \textbf{86.02}                 &                     & \textbf{81.51}                 & \textbf{88.34}                 &                     & \textbf{84.01}                 \\ \hline
\multicolumn{11}{c}{$OT$: Operating time without any scheduling algorithm (see Table~\ref{InputParameters}), $RC$: Running cost (in Indian Rupee (\rupee))}\\
\multicolumn{11}{c}{$SRT$: Scheduled Run time dictated by the developed DR algorithm, $MI$*: Multiple Intervals $\rightarrow$ illustrated in Fig.~\ref{case1b},~\ref{case2b},~\ref{case3b}}\\
\multicolumn{11}{c}{$ORC$: Optimal Running cost (in Indian Rupee (\rupee))}\\
\multicolumn{11}{c}{\textbf{Note:} In each case, $ORC$ depicts the cost of operation (including penalty) with proposed scheduling algorithm under respective pricing scheme.}\\
\multicolumn{11}{c}{\textbf{Note:} In each case, $RC$ depicts the cost of operation (including penalty) without any scheduling algorithm under respective pricing scheme.}\\
\end{tabular}
\label{AllCases}
\end{table*}

\section{Conclusion}
In this paper, a residential demand response has been demonstrated for peak  load management by load scheduling in a smart home using the proposed dynamic priority algorithm. The objective is to minimize the cost incurred to consumer herewith flattening the load profile. Preceded case studies infer that the proposed DR algorithm has shifted the time of operation of the schedulable loads present in the consumer's home as per its dynamic priority and cost to power ratio under multiple constraints. 

Three case studies have been conducted on a single consumer having same load specifications and initial conditions to reflect the robustness, efficacy, and benefits of this simple algorithm. It is observed that the consumer has dynamically entered the operating intervals of loads, which are scheduled by the algorithm in such a time--window, where the cost of electricity incurred to the consumer is minimum. This course of action has also reduced the peak demand occurred due to schedulable loads, thereby facilitated reduced peak demand penalty charges. 

This algorithm opens doors for further consideration of the operational dynamics of nonschedulable loads to develop new DR algorithms, which can predict/forecast the power consumption and usage of the nonschedulable loads, thereby including their dynamics for effective scheduling of the other loads. The type of time varying electricity tariff structure implemented in this algorithm is day ahead tariff structure, the reach of the algorithm can be extended to implement the real--time pricing based tariff structure by incorporating energy price predictions. Finally, new algorithms can push the boundaries of this algorithm by integrating it with available local renewable generation where its effective utilization and transmission of surplus energy to the grid can also be incorporated.

\bibliographystyle{IEEEtran}
\bibliography{Ref}
\end{document}